\documentclass[lettersize,journal]{IEEEtran}
\usepackage{amsmath,amsfonts}
\usepackage{algorithmic}
\usepackage{algorithm}
\usepackage{array}
\usepackage[caption=false,font=normalsize,labelfont=sf,textfont=sf]{subfig}
\usepackage{textcomp}
\usepackage{stfloats}
\usepackage{url}
\usepackage{verbatim}
\usepackage{graphicx}
\usepackage{cite}
\usepackage{comment}
\usepackage{glossaries}
\usepackage{xcolor}
\usepackage{tabularx}

\usepackage{siunitx}
\usepackage{multirow}

\newcommand{\RR}{\mathbb{R}}

\newcommand{\One}{\mathbf{1}}
\newcommand{\intd}{\, \text{d}}

\newcommand{\vm}[1]{\ensuremath{\mathbf{#1}}}

\newcommand{\transpose}{\text{T}}

\newcommand{\comments}[1]{}
\DeclareMathOperator*{\argmax}{arg\,max}

\newcommand{\pdf}{\text{pdf}}
\newcommand{\up}{\vm p} 
\newcommand{\upm}{\overline{\up}} 
\newcommand{\ups}{p} 
\newcommand{\upsm}{\overline{\ups}} 

\newcommand{\nup}{n_{\up}} 
\newcommand{\nups}{n_{\scalebox{.35}{\up}}} 
\newcommand{\dtp}{\vm x} 
\newcommand{\ndtp}{n_{\dtp}} 
\newcommand{\dtpSet}{\mathbb{X}} 

\newcommand{\Nmc}{N_{\text{MC}}} 
\newcommand{\QoI}{Q} 

\newcommand{\SD}{\Omega_{\dtp}} 

\newcommand{\tauavg}{\tau_{\text {avg}}} 
\newcommand{\taupfs}{\tau_{\text {pfs}}} 

\newacronym{pmsm}{PMSM}{permanent magnet synchronous machine}
\newacronym{im}{IM}{induction machine}
\newacronym{pms}{PM}{permanent magnets}
\newacronym{pfs}{pfs}{performance feature specifications}
\newacronym{moo}{MOO}{multi-objective optimization}
\newacronym{soo}{SOO}{single-objective optimization}
\newacronym{uq}{UQ}{uncertainty quantification}

\newacronym{pde}{PDE}{partial differential equation}
\newacronym{gpr}{GPR}{Gaussian Process Regression}
\newacronym{gp}{GP}{Gaussian Process}
\newacronym{fem}{FEM}{finite element method}
\newacronym{fe}{FE}{finite element}
\newacronym{mc}{MC}{Monte Carlo}
\newacronym{pec}{PEC}{perfect electric conductor}
\newacronym{mqs}{MQS}{magnetoquasistatic}
\newacronym{mvp}{MVP}{magnetic vector potential}
\newacronym{ga}{GA}{genetic algorithm}
\newglossaryentry{qoi}
{
	name={QoI},
	description={quantity of interest},
	first={\glsentrydesc{qoi} (\glsentrytext{qoi})},
	plural={QoIs},
	descriptionplural={quantities of interest},
	firstplural={\glsentrydescplural{qoi} (\glsentryplural{qoi})}
} 

\newglossaryentry{pm}
{
	name={PM},
	description={permanent magnet},
	first={\glsentrydesc{pm} (\glsentrytext{pm})},
	plural={PMs},
	descriptionplural={permanent magnets},
	firstplural={\glsentrydescplural{pm} (\glsentryplural{pm})}
} 

\usepackage{eso-pic}
\AddToShipoutPicture*{\footnotesize\sffamily\raisebox{0.75cm}{\hspace{1.65cm}\fbox{\parbox{\textwidth}{\copyright 2022 IEEE. Personal use of this material is permitted. Permission from IEEE must be obtained for all other uses, in any current or future media, including reprinting/republishing this material for advertising or promotional purposes, creating new collective works, for resale or redistribution to servers or lists, or reuse of any copyrighted component of this work in other works.}}}}

\begin{document}

\title{Multi-Objective Yield Optimization for Electrical Machines using Machine Learning}

\author{{Morten Christian Huber, Mona Fuhrländer$^{\star}$, Sebastian Schöps}\\ \vspace{0.2cm}
	{\small Computational Electromagnetics Group (CEM), Technische Universität Darmstadt, Germany,\\ \vspace{-0.1cm}
		$^{\star}$corresponding author's email: mona.fuhrlaender@tu-darmstadt.de}}


\maketitle

\begin{abstract}
This work deals with the design optimization of electrical machines under the consideration of manufacturing uncertainties. In order to efficiently quantify the uncertainty, blackbox machine learning methods are employed. A multi-objective optimization problem is formulated, maximizing simultaneously the reliability, i.e., the yield, and further performance objectives, e.g., the costs. A permanent magnet synchronous machine is modeled and simulated in commercial finite element simulation software. Four approaches for solving the multi-objective optimization problem are described and numerically compared, namely: $\varepsilon$-constraint scalarization, weighted sum scalarization, a multi-start weighted sum approach and a genetic algorithm. 
\end{abstract}


\section{Introduction}\label{sec:Introduction}

Electrical machines are increasingly replacing conventional combustion engines, which is an important step against climate change. However, the design and manufacturing process of electrical machines can be further improved to deal responsible with natural resources.
Computer simulations, e.g., with \gls{fem}, have the advantage that products can be virtually analyzed and optimized before the first prototypes are built and tested. This shortens the design process and saves materials.
However, typically the simulation of electrical machines is computationally very expensive, since partial differential equations have to be solved numerically, e.g., with \gls{fem}. In the approaches we present in this work, these simulations are enabled by the usage of machine learning, more precisely the usage of \gls{gpr}. The proposed workflow considers the \gls{fem} simulation tool as a blackbox such that open-source or proprietary software can be used. 

Typically, the design, i.e., the geometry and material, of a device is optimized such that performance requirements are fulfilled. However, in practice often there are manufacturing imperfections which lead to deviations in the geometry or material parameters and this may cause violations of the requirements. In this case the manufactured device is rejected. The uncertainty induced by manufacturing uncertainties can be quantified by the yield. 
The yield describes the probability, that a manufactured device fulfills all performance requirements, under consideration of uncertainties~\cite{Graeb_2007aa}.

Classic approaches for yield estimation involve \gls{mc} analysis\cite[Chap. 5]{Hammersley_1964aa}. This sampling-based approach requires the evaluation of the simulation model for each sample configuration and quickly becomes computationally prohibitive in case of \gls{fem} models of electrical machines.
Surrogate methods provide an approximation of the \gls{qoi} and proceed a \gls{mc} analysis on this approximation model, cf.~\cite{Bogoclu_2016aa}. Examples for surrogates are linear regression~\cite{Rao_1999aa}, stochastic collocation~\cite{Babuska_2007aa} and \gls{gpr}~\cite{Rasmussen_2006aa}. However, \cite[Example 3.1]{Li_2010aa} shows, that even highly accurate approximation models can lead to wrong yield estimates. For that reason, they introduced a hybrid approach, which evaluates most of the \gls{mc} sample points on the surrogate model, and a small subset of so-called critical sample points on the high-fidelity model. In~\cite{Fuhrlander_2020ab} a GPR-Hybrid approach has been recently introduced for problems in high-frequency engineering.

In order to improve the reliability of the manufacturing process, the yield shall be maximized. 
This optimization saves resources, time and money.
In practice, besides the reliability, often there are competing objectives, e.g., the minimization of material costs. 
A \gls{moo} approach is required to find a suitable trade-off between the different objectives. Classic approaches are scalarization methods, e.g., weighted sum~\cite[Chap. 3]{Ehrgott_2005aa} and $\varepsilon$-constraint~\cite[Chap. 4]{Ehrgott_2005aa}. A commonly used heuristic approach are genetic algorithms~\cite[Chap. 4]{Audet_2017aa}.

\gls{moo} under uncertainty has a long tradition in the design of electric machines \cite{Di-Barba_2010aa}. For example, the Taguchi methods, where the loss caused by uncertainty is added to the objective function as penalty term, are popular~\cite{Taguchi}. 
Another common approach is worst case optimization \cite{Bontinck_2018ae}. There, the worst case scenario is optimized, i.e., the worst performance of a design considering uncertainties. Since the objective function of the minimization problem is itself a maximization over the uncertainty range, it is a minimax problem~\cite{Ren_2013_worstcase,Graeb_2007aa}. 
For Gaussian distributed uncertainties, in six sigma design optimization the mean value of the objective function, i.e., the average performance, and the standard deviation (sigma) of the objective function, i.e., the intensity of uncertainty, are minimized, such that the constraints are fulfilled in a six sigma range around the mean value~\cite{Xiao_2015_sixsigma}.
A recent study comparing these approaches applied to the design of electrical machines is provided in~\cite{Lei2020_robust}.
In the literature, different aspects of electrical machines have been subject to optimization, e.g., the shape of the rotor~\cite{Gangl_2015topology}, the \glspl{pm}~\cite{Bontinck_2018_robust,Bontinck_2018}, or the stator slots~\cite{Arjona_CompumagNewsletter}.

In this work a \gls{pmsm}~\cite{Bontinck_2018} is modeled and simulated using CST Studio Suite\,\textregistered\,2021. Hereby, the size and position of the \glspl{pm} are considered as deterministic design parameters. The magnitude and the direction of the magnetic fields in the \glspl{pm} are considered as uncertain design parameters. The average torque is the \gls{qoi} and equipped with a lower bound it becomes a performance requirement.
We apply the GPR-Hybrid approach from~\cite{Fuhrlander_2020ab} to the modeled \gls{pmsm} in order to achieve efficient and accurate yield estimates. In contrary to~\cite{Fuhrlander_2020ab}, the uncertain parameters are not the optimization variables. This becomes especially challenging since closed-form derivatives are not available in this case. 
We state a \gls{moo} problem and provide different formulations for the different optimization methods used, i.e., for weighed sum, $\varepsilon$-constraint and the genetic algorithm. The optimization approaches are applied to the \gls{pmsm} and evaluated regarding the improvement of the objective functions and their computational efficiency.
We propose a multi-start procedure for scalarization methods, which is a heuristic globalization technique.
Eventually, we will see that the potentially prohibitive numerical effort can be reduced to an acceptable level by the employed machine learning technique.

The manuscript is structured as follows. We start with an introduction to electrical machines in general and the modeled \gls{pmsm} in particular in Sec.~\ref{sec:EM}. We continue with a section on uncertainty quantification, including the formal definition of the yield and a summary of the GPR-Hybrid approach. In Sec.~\ref{sec:MOO} the \gls{moo} problem is formulated and the different \gls{moo} techniques are introduced. In Sec.~\ref{sec:Numerics} the numerical results are presented, before we conclude with a short summary and an outlook.

\section{Electrical machine settings}\label{sec:EM}
\subsection{Brief introduction to PMSM}
As a consequence of the steadily rising focus on mitigating the origins of the climate change and the governments tightening the juridicial fundamentals to tackle it, the transport and mobility sector is currently undergoing massive changes with the change from combustion engines to electrical machines. Therefore, it is desired to deploy electrical machines with high power density and great overall efficiency. One machine type that fulfills these requirements is the \gls{pmsm}. It has several advantages over other machine types such as the \gls{im}~\cite{Melkebeek.2018}, e.g. increased overall efficiency, reduced size, high power and flux density as well as continuous operation at synchronous speed is possible.

\begin{figure}
	\centering
	\includegraphics[width=0.5\columnwidth]{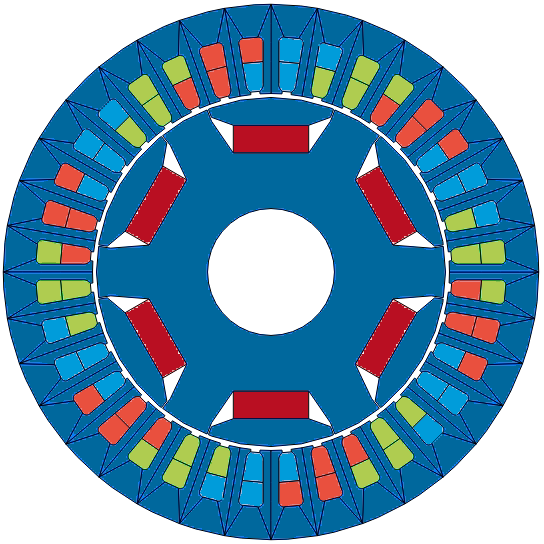}
	\caption{Design of investigated \gls{pmsm} containing three polepairs and $36$ slots.}
	\label{fig:pmsm}
\end{figure}

In the \gls{pmsm}, the rotor consists of \glspl{pm} that are placed either inside of the rotor design (interior-\gls{pmsm}) or on the surface of it (surface-\gls{pmsm}) \cite{Melkebeek.2018}. The most commonly used elements for the \gls{pm} are ceramic materials such as Barium Ferrites, rare earth materials such as Samarium and Cobalt or amorphous materials such as Neodymium Iron Boron (NdFeB) which are most often applied in the automotive area~\cite{Melkebeek.2018}. As these resources are considered limited and are subject to significant price volatilities, a minimization of the magnet dimension is desired \cite{Widmer.2015}. The stator, in contrary, is built based on a multiphase symmetrical coil winding (see Figure~\ref{fig:pmsm}) that is exciting the rotor \gls{pm}s. As the magnet dimensions of the \gls{pm} are considered to be strongly correlating with the machine costs, in this work we will minimize the magnet dimensions and thereby, we minimize the costs of manufacturing the electrical machine. The used \gls{pmsm} of this paper is based on the benchmark example introduced in \cite{DeGersem.1998}, where a manufactured \gls{pmsm} is investigated with regard to a finite element environment.

\subsection{Mathematical foundation}
The simulation is carried out based on magnetostatic formulation of Maxwell's equations since the dimensions of the problem are significantly smaller than the wave length. Therefore the displacement current density $\partial\mathbf{D}/\partial t$ is neglected. The resulting equations on a computational domain $D$ are
\begin{equation}
\nabla \times \mathbf{H} = \mathbf{J} + \nabla \times \mathbf{M},
\quad
\mathbf{B}=\mu\mathbf{H},
\quad
\mathbf{B} = \nabla \times \mathbf{A},
\label{eq:maxwell_mqs}
\end{equation}
where $\mathbf{A}$ describes the magnetic vector potential, $\mathbf{J}$ the (source) current density, $\mathbf{M}$ the magnetization, $\mu$ the permeability, $\mathbf{H}$ the magnetic field strength, $\mathbf{B}$ the magnetic field density. We suppressed spatial and temporal dependencies of the fields for brevity. The equations in \eqref{eq:maxwell_mqs} can be used to derive the \textit{curl-curl equation} that is the underlying partial differential equation to describe the phenomena of the electrical machine
\begin{equation}
\nabla \times (\nu\nabla \times \mathbf{A}) = \mathbf{J} + \nabla \times \mathbf{M},
\label{eq:curlcurl}
\end{equation} where $\nu = \mu^{-1}$ is the magnetic reluctivity. The geometry will be changed during the optimization, i.e., we parametrize the domain $D=D(\mathbf{x})$ using a parameter vector $\mathbf{x}$. On the other hand, the magnetization $\mathbf{M}:=\mathbf{M}(\up)$ will depend on uncertainties $\up$ in the magnets. Consequently, the solution $\mathbf{A}:=\mathbf{A}(\mathbf{x},\up)$ of \eqref{eq:curlcurl} inherits those dependencies. Its discretization is carried out by CST Studio Suite\,\textregistered\,2021 using higher-order \gls{fem} and a domain decomposition approach for the mechanical rotation. We address the solution for some time step $t_k$ by $\mathbf{A}_k(\mathbf{x},\up)$. 
\subsection{Assumptions} 
\label{sec:assumptions}
One of the common drivers for the usage of virtual prototypes is the event of uncertainties during the manufacturing process that influences certain machine performance measurements. These influences on the \gls{qoi} may lead to a malfunction of the whole machine. These uncertainties are often investigated in a late stage of the product development under which the efficiency of the whole product lifecycle might suffer. If those uncertainties are included in earlier design stages, more robust design development is possible as the uncertainty propagation can be estimated accurately \cite{Bontinck_2018} by means of statistical methods as it will be described in Section \ref{sec:UQ}. In the field of electrical machines, there exists a range of parameters that are potentially subject to variations in the manufacturing process~\cite{Bontinck_2018}. Multiple studies address those parameters, e.g. uncertain material properties due to the welding process, geometrical uncertainties such as rotor eccentricities or variations in the stator slots, and most importantly for the \gls{pmsm}, uncertainties that are introduced to the geometry and magnetization of the \gls{pm}s as investigated in \cite{Khan.2012, Kolb.2020, Lei.2016}. In \cite{Lei.2016}, it is shown how \gls{pm}s of three different manufacturers vary in those parameters, i.e., magnitude $B_{\mathrm{r}}$ and direction $\phi$ of the \gls{pm}'s magnetic field. To investigate their influence, we choose the average torque $\tauavg$ over one electrical period
\begin{equation}
t_{\mathrm{ep}} = \frac{1}{n \cdot N_\text{pp}} = \frac{60}{1930\,\text{rpm}\cdot 3} \approx 0.01036\,\si{\second} = 10.36\,\si{\milli\second},
\end{equation}
as the \gls{qoi}. Here, $n$ describes the speed of the machine in rpm and $N_\text{pp}$ is the number of pole pairs that are taken from the specifications in Table \ref{tab:machine_parameter}. The average torque is calculated as
\begin{equation}
\tau_{\mathrm{avg}} = \frac{1}{N_{\mathrm{per}}}\sum_{k=1}^{N_\mathrm{per}}\tau_k,
\end{equation}
where $N_\mathrm{per}$ is the number of timesteps used to resolve the electrical period in the transient simulation and $\tau_k:=\tau(\mathbf{A}_k)$, $k=1,\dots,N_\text{per}$ the value of the torque in each timestep. All model parameters are given in Table~\ref{tab:machine_parameter}.

\section{Uncertainty quantification}\label{sec:UQ}

\subsection{Definition of the yield}\label{sec:DefYield}

%
Let us distinguish between deterministic parameters $\dtp$, e.g. the geometry parameters, and uncertain parameters $\up$, e.g. the magnitude and the orientation of the magnetic field, and one or more \glspl{qoi} $\QoI(\dtp,\up)$, e.g. the torque.
The uncertain parameters are modeled as random variables. Common modeling choices are the normal or uniform distribution. Following~\cite{Bontinck_2018}, we use the uniform distribution defined by
\begin{align}
&p_i = \upsm_i + \Delta \ups_i \text{ for } i = 1,\dots,\nup \label{eq:def_uncert.para.}\\
&\text{with } \Delta \ups_i \sim \mathcal{U}(-r_i,r_i), \notag
\end{align}
where $\up = (\ups_1,\dots,\ups_{\nup})^{\transpose}$, $\upsm_i\in \RR$ and $r_i>0$. Let $\pdf(\up)$ denote the corresponding multivariate probability density function.
We introduce \gls{pfs} as constraints for the \gls{qoi}. These are the properties which should be fulfilled by the final design. 
This is given by
\begin{equation}
\QoI(\dtp,\up) \leq c,
\label{eq:pfs}
\end{equation}
keeping in mind that in case of several requirements this constraint can be extended.
Next, we introduce the safe domain as the set of uncertain parameter combinations fulfilling the \gls{pfs}, i.e.,
\begin{equation}
\SD := \left\lbrace \up: \QoI(\dtp,\up) \leq c \right\rbrace.
\label{eq:safedomain}
\end{equation}
Due to uncertainties in the manufacturing process -- represented by the random variables in~\eqref{eq:def_uncert.para.} -- it occurs that the manufactured device does not fulfill the \gls{pfs}, although the nominal design does. Following~\cite{Graeb_2007aa}, we can quantify the probability, that the manufactured device fulfills all \gls{pfs}, by the yield given by
\begin{equation}
Y(\dtp,\upm) = \int_{\RR^{\nups}} \One_{\SD} (\up) \pdf(\up) \intd \up,
\label{eq:yield}
\end{equation}
where $\One_{\SD} (\up)$ defines the indicator function with value one, if $\up$ lies in $\SD$, and zero otherwise.

\subsection{Estimation of the yield}\label{sec:EstYield}

A straightforward approach in order to estimate the yield is \gls{mc} analysis~\cite[Chap. 5]{Hammersley_1964aa}. According to the distribution of the uncertain parameters, a number of $\Nmc$ sample points $\up^{(1)},\dots,\up^{(\Nmc)}$ is generated and then the yield is estimated by
\begin{equation}
Y(\dtp,\upm) \approx Y_{\text{MC}}(\dtp,\upm) = \frac{1}{\Nmc} \sum_{i=1}^{\Nmc} \One_{\SD}(\up^{(i)}).
\label{eq:MCyield}
\end{equation}
The \gls{mc} analysis is based on the law of large numbers, which is why the standard deviation of the \gls{mc} estimator depends on the number of \gls{mc} sample points $\Nmc$, i.e.,
\begin{equation}
\sigma_{Y_{\text{MC}}} = \sqrt{\frac{Y(\dtp,\upm)(1-Y(\dtp,\upm))}{\Nmc}} \leq \frac{0.5}{\sqrt{\Nmc}}.
\label{eq:MCstd}
\end{equation}
The inequality in~\eqref{eq:MCstd} is an upper bound for the worst case scenario when $Y(\dtp,\upm)=0.5$. The $\sigma_{Y_{\text{MC}}}$ serves as measure for the accuracy of the \gls{mc} estimation and determines the required size of the \gls{mc} sample set. In the following, we aim to achieve an accuracy of $\sigma_{Y_{\text{MC}}}\leq 0.01$, which leads to a required sample size of $\Nmc = 1/(2 \cdot 0.01)^2 = 2500$.
Please note, that this sample size is well suited, 
for yield estimations until $99.9\,\%$, cf.~\cite[Chap. 4.8.7]{Graeb_2007aa}. For higher percentages ($6$-sigma) much higher accuracy is needed, i.e., more sample points or other techniques have to be applied, e.g. importance sampling~\cite{Gallimard_2019aa} or subset simulation~\cite{Bect_2017aa}.

In a \gls{mc} analysis, for each sample point, the \gls{qoi} has to be evaluated. It follows that the number of sample points required for high accuracy of the \gls{mc} analysis is prohibitive for the computational effort associated with the \gls{fe} simulations of~\eqref{eq:maxwell_mqs}. For this reason, more involved methods have been developed, e.g. importance sampling~\cite{Gallimard_2019aa}, surrogate modeling~\cite{Babuska_2007aa,Rao_1999aa,Rasmussen_2006aa} and hybrid approaches, which combine classic \gls{mc} with surrogate methods~\cite{Li_2010aa}. In this work, we extend the GPR-Hybrid approach \cite{Fuhrlander_2020ab} for efficient yield estimation to electrical machines and integrate it into the pareto optimization framework.

We briefly introduce the main ideas of the GPR-Hybrid approach. 
\gls{gpr} is a surrogate technique where the \gls{qoi} $\QoI$ is approximated by a Gaussian process. A Gaussian process can be described uniquely by its mean and covariance / kernel function. In this work we choose the mean value of the training data evaluations as mean function and the squared exponential kernel (also called radial basis function (RBF) kernel) given by
\begin{align}
\label{eq:kernel}
\text{cov}(\QoI(\vm y),\QoI(\vm y')) &= k(\vm y,\vm y')\\
&= \zeta^2 \exp{\left(-\frac{|\vm y-\vm y'|^2}{2l^2}\right)},
\nonumber
\end{align}
with the two hyperparameters $\zeta\in \RR$ and $l>0$ and two training data points $\vm y$ and $\vm y'$. Using these assumptions and a set of training data points, a surrogate model of $\QoI$ is built. Since the surrogate model is a Gaussian process, we can evaluate mean and standard deviation of the surrogate model in any (also unseen) point $\vm y^{\star}$. The mean value $\tilde{\QoI}(\vm y^{\star})$ can be used as prediction of the function value $\QoI(\vm y^{\star})$, the standard deviation $\sigma_{\text{GPR}}(\vm y^{\star})$ as an error indicator. The availability of an error indicator without additional costs, as well as 
the possibility of simple model updates due to not requiring structure in the training data set,
makes the \gls{gpr} approach so suitable for hybrid methods. For more detailed information on \gls{gpr} we refer to~\cite[Chap. 2]{Rasmussen_2006aa}.

\begin{figure}[t]%
	\centering
	{\includegraphics[width=4.cm]{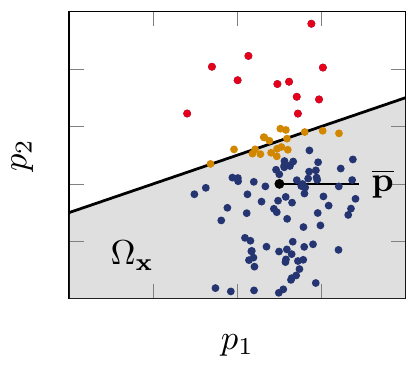}}%
	\ \ 
	{\includegraphics[width=4.cm]{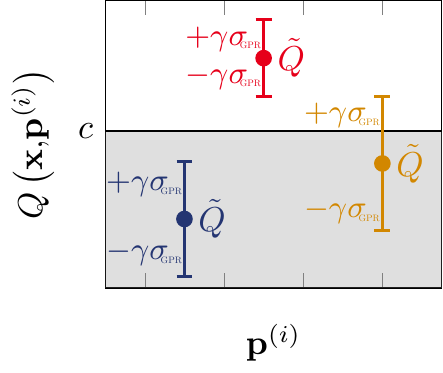}}
	\caption{Visualization of GPR-Hybrid approach with accepted (blue), not accepted (red) and critical (orange) sample points.}
	\label{fig:GPR-Hybrid}
\end{figure}

In the GPR-Hybrid approach, the training data set for the initial \gls{gpr} surrogate model is chosen small, such that the computational effort is low and the initial accuracy of the \gls{gpr} model is moderate. The \gls{mc} analysis is started, i.e., the \gls{mc} sample set is generated and step by step the sample points are evaluated with the surrogate model. Using a rule based on the prediction and the error indicator, it is decided whether the point is also evaluated on the FE model or not. Then, the sample points are classified as \textit{accepted} (lying inside the safe domain $\SD$) or \textit{not accepted}. Each time, a sample point is evaluated with the FE model, this information is used to update the \gls{gpr} surrogate model online, i.e., to increase its accuracy, and the remaining sample points are evaluated on the updated surrogate model. In the end, the classification of \textit{accepted} and \textit{not accepted} is used for the yield estimation with~\eqref{eq:MCyield}. The basic procedure is given in Algorithm~\ref{algo:GPR-Hybrid} and a visualization of the classification of the sample points is shown in Fig.~\ref{fig:GPR-Hybrid}. In this work, the safety factor is set to $\gamma = 2$. Please note that higher values of the safety factor may increase the number of FE model evaluations, but promise higher accuracy of the yield estimator. For more details we refer to~\cite{Fuhrlander_2020ab}.
\begin{algorithm}[t]
	\caption{GPR-Hybrid approach}
	\begin{algorithmic}[1]
		\STATE{\textbf{Input:} \gls{mc} sample points $\up^{(i)}$, $i=1,\dots,\Nmc$, deterministic design point $\dtp$, \gls{pfs} threshold $c$, initial GPR surrogate model $\tilde{\QoI}$}, 
		\FOR{$i=1,\dots,\Nmc$}
		\STATE{Evaluate the GPR surrogate model and obtain $\tilde{\QoI}(\dtp,\up^{(i)})$ and $\sigma_{\text{GPR}}(\dtp,\up^{(i)})$}
		\IF{$\tilde{\QoI}(\dtp,\up^{(i)}) + \gamma \sigma_{\text{GPR}}(\dtp,\up^{(i)}) \leq c $}
		\STATE{Classify $\up^{(i)}\in \SD$ (accepted)}
		\ELSIF{$\tilde{\QoI}(\dtp,\up^{(i)}) - \gamma \sigma_{\text{GPR}}(\dtp,\up^{(i)}) \geq c $}
		\STATE{Classify $\up^{(i)}\notin \SD$ (not accepted)}
		\ELSE
		\STATE{$\up^{(i)}$ is a critical sample point}
		\STATE{Evaluate the FE model and obtain $\QoI(\dtp,\up^{(i)})$\\
			\IF{$\QoI(\dtp,\up^{(i)}) \leq c$ }
			\STATE{Classify $\up^{(i)}\in \SD$ (accepted)}
			\ELSE
			\STATE{Classify $\up^{(i)}\notin \SD$ (not accepted)}
			\ENDIF \\
			Update GPR surrogate model $\tilde{\QoI}$	}
		\ENDIF
		\ENDFOR
	\end{algorithmic}
	\label{algo:GPR-Hybrid}
\end{algorithm}

\section{The multi-objective optimization problem}\label{sec:MOO}

\subsection{Formulation of the optimization problem}\label{sec:MOO_fromulation}

Having an efficient method for yield estimation, we formulate the optimization problem for the maximization of the yield, i.e., the maximization of the probability that one realization in a manufacturing process fulfills all performance feature specifications. 
As optimization variables, the uncertain design parameters can be considered as for example in~\cite{Fuhrlander_2021aa}, or both, uncertain and deterministic design parameters can be considered as in~\cite{Fuhrlander_2021ab}.
In this work we will focus on deterministic design parameters as optimization variables, i.e., the optimization problem reads
\begin{equation}
\max_{\dtp\in \dtpSet} Y(\dtp,\upm) \label{eq:SOO},
\end{equation}
where $\dtpSet \subseteq \RR^{\ndtp}$ is the feasible set.
In practice, often there is not only the reliability which has to be optimized during the design process, but there are further objectives. This leads to a \gls{moo} setting
\begin{align}
&\max_{\dtp\in \dtpSet} f_1(\dtp,\upm) \equiv Y(\dtp,\upm) \label{eq:MOO} \\
&\max_{\dtp\in \dtpSet} f_i(\dtp,\upm), \ \ i=2,\dots,k \notag
\end{align}
with $k$ objective functions.

\begin{figure}[t]%
	\centering
	{\includegraphics[width=0.25\textwidth]{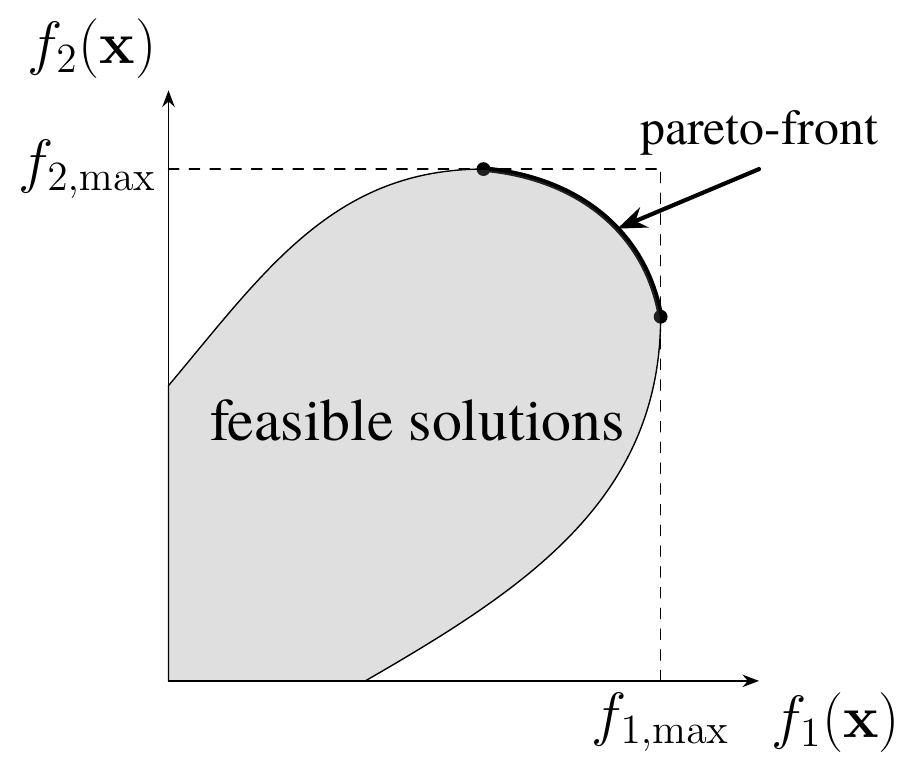}}
	\caption{Visualization of the pareto-front for a bi-objective maximization problem.}
	\label{fig:ParetoFront}
\end{figure}

While in \gls{soo} the aim is to find the best solution with respect to \textit{one} objective function, in \gls{moo} the aim is to find the best solution with respect to \textit{all} objective functions. Since the best solution for the first objective function is, in general, not the best solution for the second objective function, the concept of so-called pareto-optimal solutions have been introduced~\cite[Def. 2.1]{Ehrgott_2005aa}: A feasible solution $\dtp^{\star} \in \dtpSet$ is pareto-optimal, if there is no $\dtp \in \dtpSet$ such that 
\begin{align}
&f_i(\dtp) \geq f_i(\dtp^{\star}) \text{ for } i=1,\dots,k \text { and } \label{eq:Def_pareto}\\
&f_i(\dtp)>f_i(\dtp^{\star}) \text{ for some } i\in\lbrace 1,\dots,k\rbrace. \notag
\end{align}
This implies that a pareto-optimal solution is a solution, such that it is not possible to improve the value of one objective function without deteriorating the value of another one.
Thus, the solution of a \gls{moo} problem is not one single point, but the set of pareto-optimal solutions, the so-called pareto-front. In Fig.~\ref{fig:ParetoFront} the pareto-front of a bi-objective maximization problem is visualized.

\subsection{Scalarization methods}\label{sec:MOO_scalarization}

A classic approach for \gls{moo} are scalarization techniques like the weighted sum or $\varepsilon$-constraint methods~\cite[Chap. 3-4]{Ehrgott_2005aa}. The basic idea of scalarization is to transform the \gls{moo} problem into a \gls{soo} problem. In case of the weighted sum scalarization this is achieved by combining all objective functions into one weighted sum which is the new objective function. We obtain
\begin{equation}
\max_{\dtp\in \dtpSet} w_{1}Y(\dtp,\upm) + \sum_{i=2}^{k} w_i f_i(\dtp,\upm) \label{eq:MOO_weightedSum}
\end{equation}
with weights $w_i> 0$, $i=1,\dots,k$. In case of the $\varepsilon$-constraint scalarization, only one of the objective functions is maintained, while the remaining objective functions are included into the constraints. This leads for example to the following formulation
\begin{align}
&\max_{\dtp\in \dtpSet} Y(\dtp,\upm) \label{eq:MOO_epsConstraint} \\
&\text{ s.t. } f_i(\dtp,\upm) \geq \varepsilon_{i}, \ \ i=2,\dots,k \notag
\end{align}
with lower bounds $\varepsilon_{i}\in \RR$, $i=2,\dots,k$. Please note that $k$ different formulations are possible, depending on the choice of the objective function which stays an objective function.
The advantage of scalarization techniques is the fact, that after the reformulation we have a \gls{soo} problem, which can be solved with classic \gls{soo} algorithms like for example simplex algorithms~\cite{Powell_2007aa} or Sequential Quadratic Programming~\cite[Chap. 19]{Ulbrich_2012aa}.
On the other hand, one disadvantage of scalarization is that you obtain only one pareto-optimal solution (not a pareto-optimal front), and this solution depends on the choice of the weights $w_i$ or the bounds $\varepsilon_{i}$, respectively. Thus, in order to approximate the pareto-front, the reformulated problem has to be solved several times with different values for weights and bounds.
Another disadvantage of the weighted sum method is, that convexity requirements on the feasible set $\dtpSet$ are required, in order to have the chance to find each pareto-optimal solution. For more details on that we refer to~\cite[Chap.~3]{Ehrgott_2005aa}.

\subsection{Multi-start procedure}\label{sec:MOO_multitstart}

When using the weighted sum method, we will also investigate a multi-start procedure. By this we try to cover the feasible set best without being trapped in a local optimum. Therefor, a set of starting points is generated and for each starting point the optimization is started. Inspired by the terms of reinforcement learning~\cite{Sutton_1998aa}, we refer to this phase as exploration phase, since the aim is to explore as much of the feasible set as possible in order to find the most promising \textit{region}. After a fixed (not to high) number of objective function evaluations or iterations, the results are compared. Then, the optimization is continued with the solution which had the best objective function value after the exploration phase. This second phase we refer to as the exploitation phase, because now we seek for the best value in this region. 

Multi-start procedures are a commonly used heuristic for globalizing local optimization solvers~\cite[Chap. 6]{Marti_2018aa} and appear in many different forms. In its simplest form, several starting points are used, the local optimization is run till the end and then the best solution is chosen. In our setting, each evaluation of the objective function is computationally expensive. Thus, running the local optimization multiple times might be prohibitive. For that reason we establish the two phases and follow only the most promising solution from the exploration phase. The computing time can be further improved, by using lower fidelity models for the exploration phase and a high fidelity model in the exploitation phase. There are many options to distinguish between low and high fidelity models: when using \gls{fem} it can refer to the number of elements for example, or in case of yield optimization, it can refer to the size of the \gls{mc} sample set. The basis procedure is sketched in Algorithm~\ref{algo:Multi-start}.
\begin{algorithm}[t]
	\caption{Multi-start optimization method}
	\begin{algorithmic}[1]
		\STATE{\textbf{Input:} set of starting points $\dtp_0^{(i)}$, $i,\dots,N_{\text{s}}$, low fidelity model $f_{\text{L}}$, high fidelity model $f_{\text{H}}$, optimization problem (P)}, 
		\FOR{$i=1,\dots,N_{\text{s}}$}
		\STATE{Solve (P) based on $f_{\text{L}}$, starting with $\dtp_0^{(i)}$ till stopping criterion for exploration phase is reached}
		\STATE{Obtain best solution $\dtp_{\star}^{(i)}$}
		\ENDFOR
		\STATE{Set $\dtp_{\text{continue}} = \argmax_{i=1,\dots,N_{\text{s}}} f_{\text{L}}(\dtp_{\star}^{(i)})$}
		\STATE{Solve (P) based on $f_{\text{H}}$, starting with $\dtp_{\text{continue}}$ till stopping criterion for exploitation phase is reached}
	\end{algorithmic}
	\label{algo:Multi-start}
\end{algorithm}

\subsection{Genetic algorithms}\label{sec:MOO_genetic}

An alternative to scalarization methods are genetic algorithms~\cite[Chap. 4]{Audet_2017aa}. They, in fact, solve \gls{moo} problems directly and the solution is an approximation of the pareto-front.
Genetic algorithms are motivated by the process of natural evolution of organisms. 
The basic idea is as follows: a so-called population of individuals, i.e., an initial set of data points, is generated, the objective functions are evaluated on these individuals and in each iteration the population is updated. The update consists of three main elements: selection (survival of the fittest), crossover (reproduction process where genetic traits of the individuals are propagated) and mutation (variation of genetic traits). The rules how these elements are applied, depend on the so-called fitness, a metric for the quality of an individual based on the objective function values.

\section{Numerical results}\label{sec:Numerics}

\subsection{Problem setting}
\begin{table}[bt]
	\centering
	\caption{Overview over most important machine parameters~\cite{Pahner_1998aa,Bontinck_2018}.}
	\begin{tabular}{c  c}
		\hline 
		\hline
		Parameter name & Value\\
		\hline
		Rotor rated speed $n$ & $1930$\,$\mathrm{rpm}$  \\
		Rated current $I_{\mathrm{eff}}$	& \SI{10.60}{\ampere}\\
		Average torque (sim.) $\tau_{\mathrm{avg}}$ & \SI{10.64}{\si{\newton\meter}}\\
		Windings per coil $N_{\mathrm{c}}$ & \SI{12}{}\\
		Stator windings in series $a$ & \SI{1}{}\\
		Number of pole pairs $N_\text{pp}$& \SI{3}{}\\
		Number of slots $Q$ & \SI{36}{}\\
		Number of phases $m$ & \SI{3}{}\\
		Number of slots per pole and phase $q$ & \SI{2}{}\\
		Winding pitch $W/\tau_\mathrm{p}$ & $5/6$\\
		\hline 
		\hline 
	\end{tabular}
	\label{tab:machine_parameter}
\end{table}
\begin{figure}[b]
	\centering
	\includegraphics[width=0.4\columnwidth]{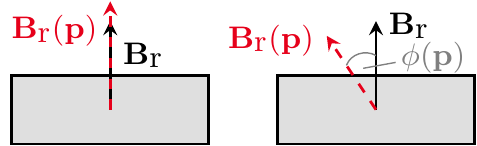}
	\caption{Geometrical depiction of uncertain rotor design parameters based on \cite{Bontinck_2018}.}
	\label{fig:uncert_design}
\end{figure}
\begin{figure}[t]
	\centering
	\includegraphics[width=0.6\columnwidth]{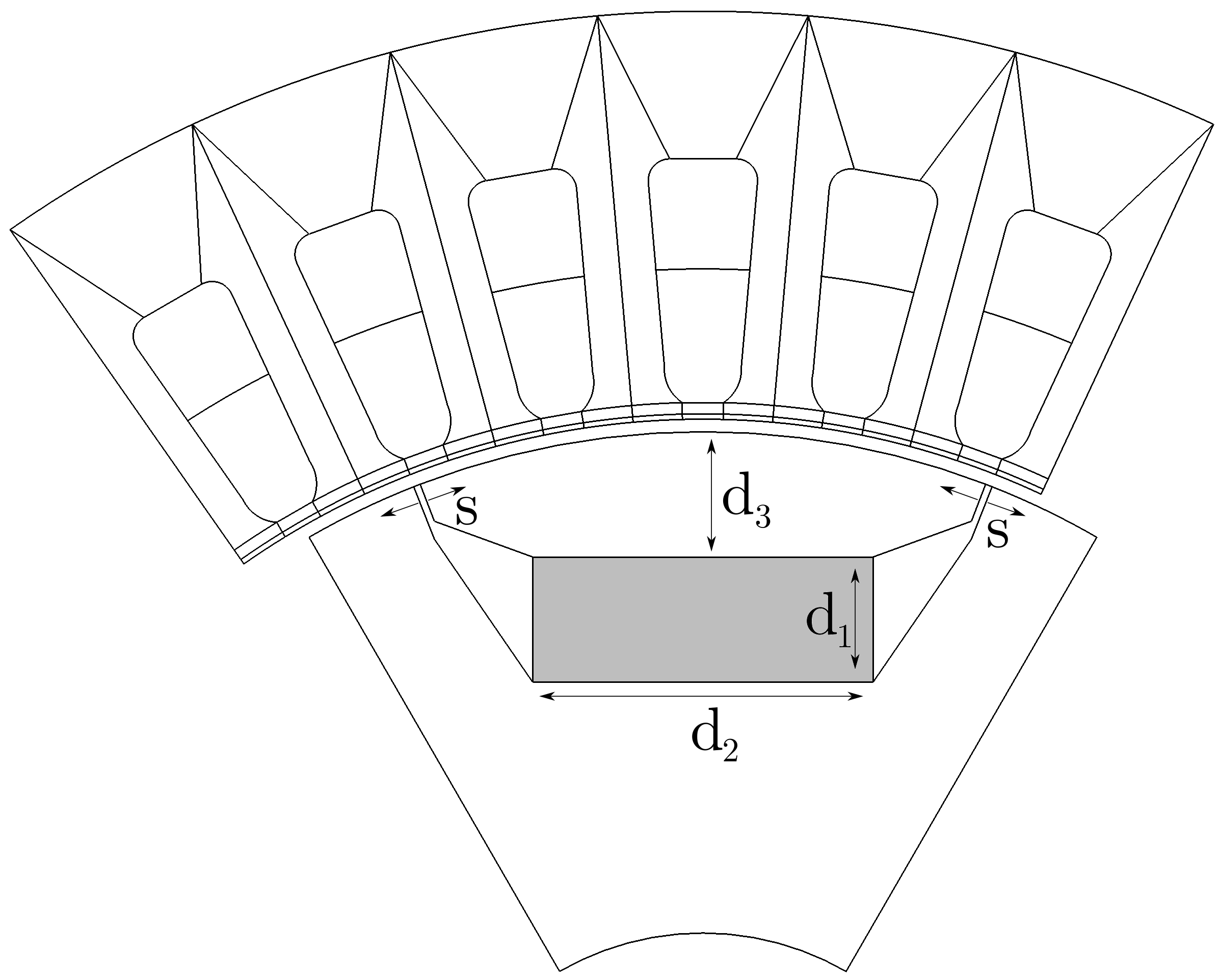}
	\caption{Geometrical depiction of deterministic motor design parameters $\mathbf{x}$.}
	\label{fig:det_design}
\end{figure}
The machine under study is the \gls{pmsm} from \cite{DeGersem.1998, Bontinck_2018} with the properties as specified in Table \ref{tab:machine_parameter}. For the uncertainty quantification the deterministic and uncertain parameters need to be concretized. According to measurement studies of a manufactured \gls{pmsm} in \cite{Hameyer.2013}, the magnitude ${B}_{\text{r},i}$ and the direction ${\phi}_i$ of the magnetic field of the $i$-th permanent magnet are consequently assumed uncertain, i.e., they are modeled as uniformly distributed random variables with mean values 
\begin{equation}
\begin{split}
\upm &= \left(\overline{B}_{\text{r},1}, \dots,  \overline{B}_{\text{r},6}, \overline{\phi}_1, \dots, \overline{\phi}_6 \right)\\
&= \left(\SI{0.94}{\tesla}, \dots, \SI{0.94}{\tesla},\SI{0}{\degree}, \dots, \SI{0}{\degree} \right)
\end{split}
\end{equation}(see Fig. \ref{fig:uncert_design}). According to Eq. \eqref{eq:def_uncert.para.}, we write
\begin{align}
&p_i = \upsm_i + \Delta \ups_i \text{ for } i = 1,\dots,12 \label{eq:def_uncert.para_concrete}\\
&\text{with } \Delta \ups_{i} \sim \mathcal{U}(\SI{-0.05}{\tesla},\SI{0.05}{\tesla}) \text{ for } i = 1,\dots,6 \notag\\
&\text{and } \Delta \ups_{i} \sim \mathcal{U}(\SI{-3}{\degree},\SI{3}{\degree})\text{ for } i = 7,\dots,12. \notag\\
\end{align}
As the deterministic design parameters we consider the geometric parameters describing the rotor geometry as depicted in Fig.~\ref{fig:det_design}, i.e., 
\begin{equation}
\begin{split}
\mathbf{x} &= \left( d_1,d_2,d_3,s \right)\\ 
&=\left(\SI{19}{\milli\meter}, \SI{7}{\milli\meter}, \SI{7}{\milli\meter}, \SI{0}{\degree}\right).
\end{split}		
\end{equation} 
For the mathematical problem formulation of the electrical machine, the magnetostatic simplification of Maxwell's equations \eqref{eq:maxwell_mqs} is used. The discretization is carried out inside of CST Studio Suite\,\textregistered\,2021 with \gls{fem} to acquire a solution for $\mathbf{A}$ in the computational domain (later referred to as \gls{fe} model). This allows to calculate the \gls{qoi}, i.e., the average torque $\tauavg$ over one electrical period. Next, the \gls{qoi} is combined with a \gls{pfs} that includes a lower bound $\taupfs$  for the average torque~$\tauavg$
\begin{equation}
Q(\mathbf{x},\mathbf{p}) := \tauavg\left(\mathbf{A}(\mathbf{x},\mathbf{p})\right) \geq \taupfs.
\label{eq:qoi}
\end{equation}
Now we have everything at hand in order to estimate the yield based on (\ref{eq:safedomain}--\ref{eq:yield}).
The general problem formulation for the optimization is based on \cite[Eq. (9.52)]{Bontinck_2018} and is modified for the purpose of a multi-objective yield and size, i.e., cost, optimization.
The modified problem formulation is given by
\begin{align}
\max_{\vm x \in \RR^4} & \hspace{0.2cm} Y(\dtp,\upm) &&\approx Y_{\text{MC}}(\dtp,\upm) = \frac{1}{\Nmc} \sum_{i=1}^{\Nmc} \One_{\SD}(\up^{(i)})  \label{eq:objfunc_yield}\\
\min_{\vm x \in \RR^4} &\hspace{0.2cm}C(\vm x) &&:= d_1\,d_2 \label{eq:objfunc_cost}\\
\text{s.t.} &\hspace{0.1cm}\dtp &&\geq \dtp_{\mathrm{lb}}   \label{eq:constraint1} \\
&\hspace{0.2cm} d_3&&\leq d_{3,\mathrm{ub}}  \label{eq:constraint2} \\
&\hspace{0.2cm} d_2+d_3&& \leq 15   \label{eq:constraint3} \\
&\hspace{0.2cm} 3d_1 -2d_3 &&\leq 50   \label{eq:constraint4} \\
&\hspace{0.2cm} s &&\leq s_{\mathrm{ub}},   \label{eq:constraint5}
\end{align}
where $C(\mathbf{x})$ denotes the additional quantity of the \gls{pm} surface dimensions that directly correlate with the costs of the \gls{pm}. As only the surface dimension is varied during the optimization routine and the magnet depth is not considered within the optimization, the surface dimension is $\SI{}{\milli\meter\squared}$ and consequently considered proportional to the costs.\par 
The stated problem will be tackled with the approaches introduced in Sec. \ref{sec:MOO}. In the applied scalarization methods, the \gls{moo} is transformed into a \gls{soo}. 
The \gls{soo} framework is built in python using the PDFO framework \cite{ZhangTomM.RagonneauandZaikun.20.09.2021} by M.J.D. Powell that contains optimization methods to efficiently solve \gls{soo} problems with linear or nonlinear constraints based on simplex and trust-region methods (see \cite{Powell.1998, Powell.2014, Powell_2007aa} for more details). For the nonlinearly constrained \gls{soo} problem COBYLA will be used, which stands for Constrained Optimization By Linear Approximation. The problem formulation including linear constraints will be solved with LINCOA, which stands for Linearly Constrained Optimization Algorithm. Finally, the Genetic Algorithm simulation will be carried out with the pymoo framework of python \cite{pymoo} and uses the NSGA-II algorithm \cite{nsga_2}.
\subsection{Training of GPR model}
In this section the number of training data points for the initial \gls{gpr} model is derived and the underlying distribution is explained. The two investigated sets of training sample points are drawn from uniform distributions that, first (Case $\#1$), vary in all parameters (features) $\mathbf{x}, \upm$ and second (Case $\#2$), only in the mean of the uncertain parameters $\upm$. One shoud note that the latter case describes the sample point distributions as performed in the later \gls{mc} analysis to conduct an uncertainty quantification for the initial set of deterministic parameters, whereas the first case describes the process that occurs within the optimization where the deterministic parameters are used as the optimization variables, and, are consequently varied throughout the optimization iterations. The reference yield estimate for $\Nmc=2500$ as derived in Sec. \ref{sec:EstYield} that is achieved by solely simulating the \gls{mc} analysis on the FE model with training sample points as in Case $\#1$ is  $Y_{\text{MC}}^{\text{ (CST)}}(\dtp,\upm) = 0.0428$. With the \gls{mc} error indicator from \eqref{eq:MCstd} we derive the lower and upper bound of valid yield predictions to use for the validation of the proposed \gls{gpr}-Hybrid method. We get
\begin{equation}
\sigma_{Y_{\text{MC}}} = \sqrt{\frac{Y_{\text{MC}}^{\text{ (CST)}}(\dtp,\upm)(1-Y_{\text{MC}}^{\text{ (CST)}}(\dtp,\upm))}{\Nmc}} \approx 0.00405.
\end{equation}
Consequently, all predictions in the $\sigma_{Y_{\text{MC}}}$ region of the yield estimate $Y_{\text{MC}}^{\text{(CST)}}$, and therefore in $[0.03875, 0.04685]$, are assumed as valid. In the following $\tilde{Y}_\mathrm{MC}(\dtp,\upm)$ describes the yield estimation achieved with the \gls{gpr}-Hybrid method. Table~\ref{tab:gprtrain} summarizes the simulation results for the \gls{gpr}-Hybrid approach for the two proposed cases from above. Here, $N_\mathrm{train}$ describes the number of training sample points used (offline), $N_\mathrm{online}$ is the number of re-evaluations on the FE model (online) within the \gls{gpr}-Hybrid approach, $N_\mathrm{GPR}$ is the number of evaluations on the \gls{gpr} model and $N_\mathrm{train,tot}$ is the sum of offline and online usage of sample points, i.e., $N_\mathrm{tot} = N_\mathrm{train} + N_\mathrm{online}$.
\begin{table}[b]
	\centering
	\renewcommand{\arraystretch}{1.4}
	\caption{Overview over yield estimations achieved with the \gls{gpr}-Hybrid approach based on the two proposed cases of training sample points.}
	\begin{tabular}{l | l l l l l}
		\hline
		\hline
		& $N_\mathrm{train}$ & $\tilde{Y}_\mathrm{MC}(\dtp,\upm)$ & $N_\mathrm{online}$ & $N_\mathrm{GPR}$ & $N_\mathrm{tot}$\\
		\hline
		\multirow{3}{*}{Case $\#1$}  & $10$ &$0.0428$ & $26$ & $2474$ & $10+26=36$\\
		& $20$ &$0.0428$ & $23$ & $2477$ &$20+23=43$\\
		& $50$ &$0.0428$ & $21$ & $2479$ &$50+21=71$\\
		\hline 
		\multirow{3}{*}{Case $\#2$}  & $15$ &$0.0392$ & $34$ & $2466$ &$15+34 =49$\\
		& $20$ &$0.0436$ & $40$ & $2460$ &$20+40 =60$\\
		& $30$ & $0.0432$ & $43$ &$2457$&$30+43=73$\\
		\hline 
		\hline
	\end{tabular}
	\label{tab:gprtrain}
\end{table}

\begin{figure}
	\centering\includegraphics[width=\columnwidth]{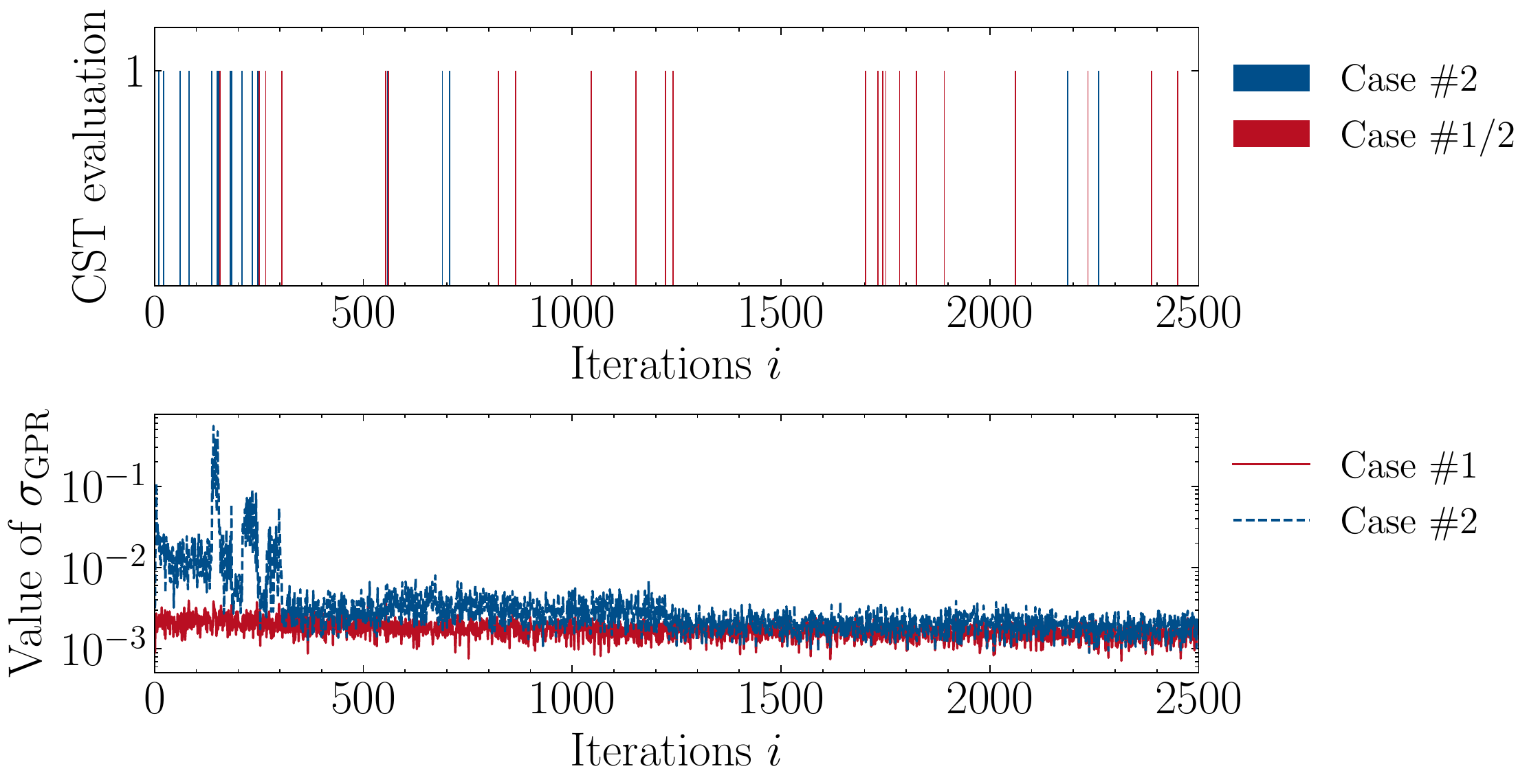}
	\caption{Comparison of FE model evaluations and \gls{gpr} standard deviation $\sigma_{\text{GPR}}$ propagation for the two proposed test cases. Iteration $i$ refers to th $i$-th \gls{mc} sample point within the \gls{gpr}-Hybrid approach.}
	\label{fig:comparison_training}
\end{figure}
It can be seen that in Case~$\#1$ the GPR is capable of delivering very accurate predictions with only $10$ initial training data points. The same holds for all considered training set sizes above $10$. 
On the other hand, the GPR built in Case~$\#2$, requires a higher number of training data points for accurate predictions -- at least $15$. 
The deviations in the yield estimate lie all inside the accepted error tolerance determined by $\Nmc=2500$.
It is evident that this prediction performance is a bit worse when compared to Case $\#1$ as the training set does not solely include points that resemble the sample points drawn in a \gls{mc} analysis, but also the deterministic optimization variables. However, as in the remainder of this paper an optimization over the deterministic parameters is desired, a variation in the deterministic parameters $\mathbf{x}$ is used in the training. Figure \ref{fig:comparison_training} shows the iterations in which a re-evaluation on the original model was needed and the correlation to the standard deviation $\sigma_{\text{GPR}}$ of the \gls{gpr} model prediction. Here, Case $\#1/2$ describes the iterations in which a re-evaluation was needed for both cases. It can be observed that in Case~$\#2$ more FE model evaluations are needed in the early stages of the \gls{mc} analysis that lead to a significant reduction of the standard deviation $\sigma_{\text{GPR}}$ after approximately $250$ iterations converging to the value of the standard deviation that was achieved in Case~$\#1$.
With $20$ training data points in Case~$\#2$ we have total costs of $60$ objective function evaluations and only two sample point differently classified as in the reference solution. In all further experiments we use this size of initial training data points for the optimization.
With this, we obtain an approximation not only in the starting point, but in a neighborhood of the starting point. Furthermore, it has to be noted that with the applied \gls{gpr}-Hybrid approach the number of needed FE model evaluations that take around $t_\mathrm{FE}=\SI{85}{\second}$ is significantly reduced.

\subsection{Optimization results}
The parameters of both the $\varepsilon$-constraint and the weighted sum method are varied throughout various optimizations. Additionally, a multi-start procedure is utilized. Finally, a genetic algorithm from the python pymoo package is used and applied to the individual objectives.
\subsubsection{$\varepsilon$-constraint method}
\begin{figure}[t!]
	\centering
	\includegraphics[width=\columnwidth]{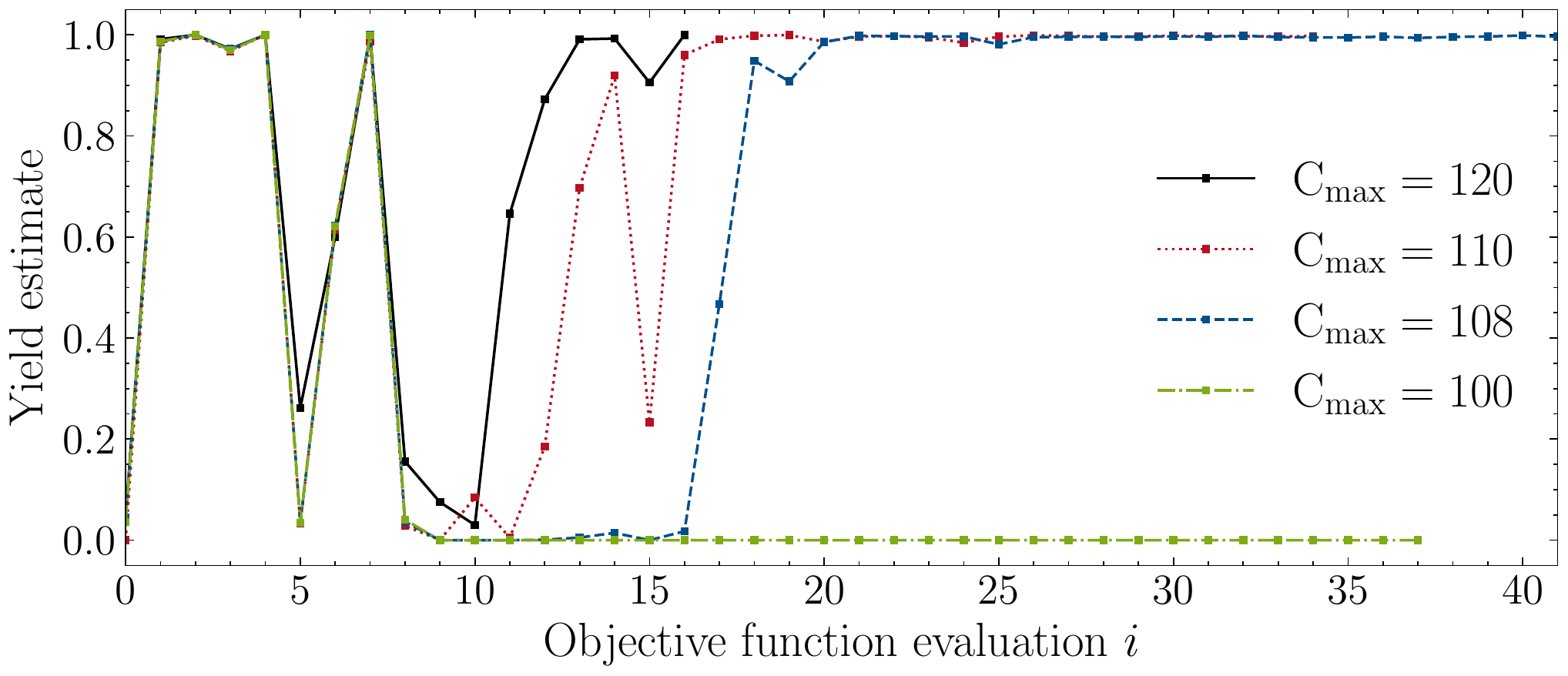}
	\caption{Propagation of the yield estimate throughout the optimization for the $\varepsilon$-constraint method.}
	\label{fig:con_yield}
\end{figure}

\begin{figure}[t!]
	\centering
	\includegraphics[width=\columnwidth]{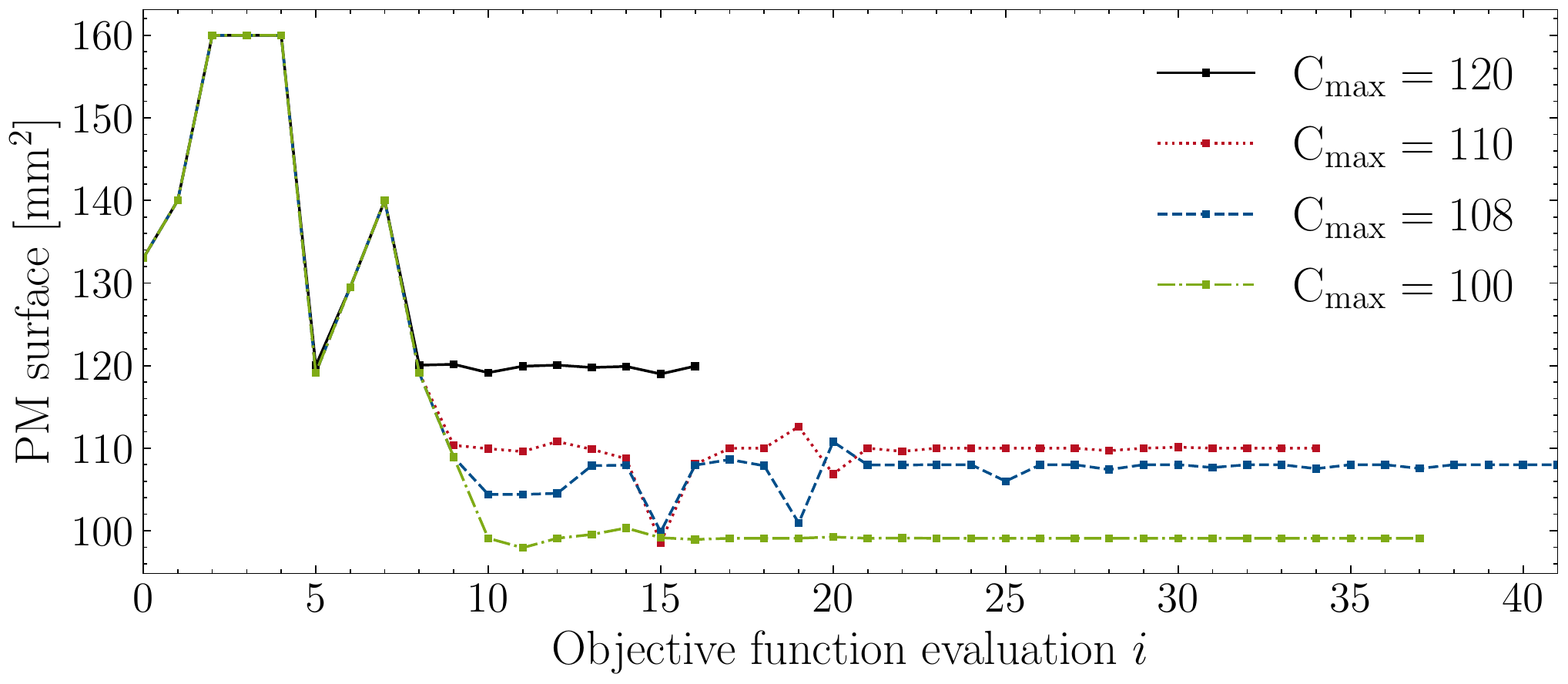}
	\caption{Propagation of the PM surface dimensions throughout the optimization for the $\varepsilon$-constraint method.}
	\label{fig:con_cost}
\end{figure}

\begin{table}[b]
	\begin{center}
		\caption{Computation statistics of $\varepsilon$-constraint method with variation in $C_\mathrm{max}$.}
		\label{tab:stats_con}
		\def\arraystretch{1.2}
		\begin{tabular}{l|l|l|l|l}
			\hline
			\hline
			$C_\mathrm{max}$ &$ \tilde{Y}_\mathrm{MC}^{(\mathrm{opt})}$ & $C(\mathbf{x})$ &   $n_\mathrm{fev}$ &   Total FE evaluations \\
			\hline
			$120$  &$1.0$ & $119.938$ & $17$ &   $20 + 350 = 370$ \\
			$110$ &  $0.998$ & $110.000$ & $35$ &  $20 +362 = 382$ \\
			$108$  &$0.996$ & $108.000$ & $42$ &  $20 + 315 = 335$\\
			$100$  &$0.0$ & $99.092$ & $38$ & $20 + 162 = 182$\\
			\hline  
			\hline
		\end{tabular}
	\end{center}
\end{table}
As introduced in Eq. \eqref{eq:MOO_epsConstraint}, one of the objective functions is transformed into a constraint. The yield function remains objective, while the function of the surface of the \gls{pm} becomes a constraint, bounded by $\varepsilon~\equiv~C_\mathrm{max}$. The resulting optimization problem reads
\begin{align}
\max_{\vm x \in \RR^4} & \hspace{0.2cm} Y(\dtp,\upm) \label{eq:Numerics_MOO_constraint}\\
\text{s.t.} &\hspace{0.1cm} C(\dtp)  \leq C_\mathrm{max} \notag\\
& \hspace{0.1cm} \text{\eqref{eq:constraint1} -- \eqref{eq:constraint5} hold.} \notag
\end{align}
As outlined above, the optimization problem is solved with the COBYLA solver of the PDFO framework. The outcome of the simulation can be observed in Table \ref{tab:stats_con} that shows the chosen bound $C_\mathrm{max}$, the optimized yield estimate $ \tilde{Y}_\mathrm{MC}^{(\mathrm{opt})}$, the \gls{pm} surface dimensions of the optimized motor design $C(\mathbf{x})$, the number of objective function calls $n_\mathrm{fev}$ during the optimization and the number of needed FE model evaluations given as sum of offline and online evaluations. The utilized solver is capable of significantly increasing the yield $\tilde{Y}_\mathrm{MC}$ for multiple choices of $C_\mathrm{max}$ and the resulting surface dimensions are always very close to the upper limit $C_\mathrm{max}$. But it can additionally be observed that if the value of $C_\mathrm{max}$ is chosen too low, the optimization routine fails to improve the yield estimate ($C_\mathrm{max}=100$). We see that the choice of $\varepsilon=C_\mathrm{max}$ is a trade off between the optimality of the yield and the surface dimensions and is crucial for the result of the optimization, which poses a drawback in practice. The propagation of the yield estimate and the surface dimensions can be seen in Figure \ref{fig:con_yield} and \ref{fig:con_cost}. The exploration phase of the simplex methods is observable in the first ten function evaluations as the results for all different settings look comparable in that range. Furthermore, it is evident that the $\varepsilon$-constraint method is varying the design parameters $\mathbf{x}$ such that an improvement of the yield estimate is achieved in a few function evaluations only (e.g., for $C_\mathrm{max}=108$ between the 16th and 18th function evaluation in Figure \ref{fig:con_yield}). The same phenomenon is visible in Figure \ref{fig:con_cost} for the \gls{pm} surface dimensions, where the desired upper bouadary for the \gls{pm} surface dimensions $C_\mathrm{max}$ are quickly met. The termination of the algorithm is determined by the applied simplex methods and the correlating trust region subproblem (see \cite{Powell.1998, Powell.2014, Powell_2007aa} for more details).

\subsubsection{Weighted sum method}
\begin{table}[b]
	\begin{center}
		\caption{Computation statistics of weighted sum method with variation in the weight $w$.}
		\label{tab:stats_ws}
		\def\arraystretch{1.2}
		\begin{tabular}{l|l|l|l|l}
			\hline
			\hline
			$w$ &$ \tilde{Y}_\mathrm{MC}^{(\mathrm{opt})}$ & $C(\mathbf{x})$ &   $n_\mathrm{fev}$ &   Total FE evaluations \\
			\hline
			$1\times10^{-3}$  &$0.999$ & $128.963$ & $31$ &   $20 + 259 = 279$ \\
			$2\times10^{-3}$ &  $0.996$ & $114.594$ & $35$ &  $20 +230 = 250$ \\
			$3\times10^{-3}$  &$0.995$ & $109.7022$ & $59$ &  $20 + 262 = 282$\\
			$5\times10^{-3}$  &$0.978$ & $105.174$ & $100$ &  $20 + 734 = 754$\\
			\hline  
			\hline
		\end{tabular}
	\end{center}
\end{table}
In contrary to the $\varepsilon$-constraint method, the weighted sum method combines both objectives into a weighted sum as in Equation \eqref{eq:MOO_weightedSum} that is consequently minimized. 
The resulting problem formulation reads
\begin{align}
\min_{\vm x \in \RR^4} & \hspace{0.2cm} f(\dtp,\upm) = -Y(\dtp,\upm) + w\,C(\dtp) \label{eq:Numerics_MOO_sum}\\
\text{s.t.} &\hspace{0.2cm} \text{\eqref{eq:constraint1} -- \eqref{eq:constraint5} hold.} \notag
\end{align}
Therefore, the degree of freedom in that formulation is the weight $w$ that is varied throughout the optimizations conducted with the weighted sum method.
The optimization problem is solved with the LINCOA solver of the PDFO framework. The outcome is given in Table \ref{tab:stats_ws}. The weighted sum approach also proves to significantly increase the yield estimate $\tilde{Y}_\mathrm{MC}$ of the optimization problem in Eq. (\ref{eq:objfunc_yield}--\ref{eq:constraint5}). The weight $w$ is varied from $1\times10^{-3}$ to 	$5\times10^{-3}$ in four steps. It is evident that the simulation needs more function evaluations when the weight $w$ is increased and, consequently, the surface dimensions $C(\mathbf{x})$ have an increased influence on the objective function (e.g., $n_\mathrm{fev}=31$ for $1\times10^{-3}$ and $n_\mathrm{fev}=100$ for $5\times10^{-3}$). In the latter case the simulation is terminated because it reached the upper limit of $n_\mathrm{fev}=100$ function evaluations. Similarily to the COBYLA, the LINCOA terminates once the trust region conditions are fulfilled and additionally takes into account the weighted difference of the last four function evaluations given by 
\begin{equation}
\Delta_{f_j}=\left|\left(\frac{1}{4} \sum_{i=1}^{4} f_{j-i}(\mathbf{x},\up)\right) - f_j(\mathbf{x},\up)\right|,
\label{eq:algo_termination}
\end{equation} 
where ${f_j}(\mathbf{x,\up})$ is the objective function value of the $j$-th function evaluation. 
\par
\begin{figure}[t!]
	\centering
	\includegraphics[width=\columnwidth]{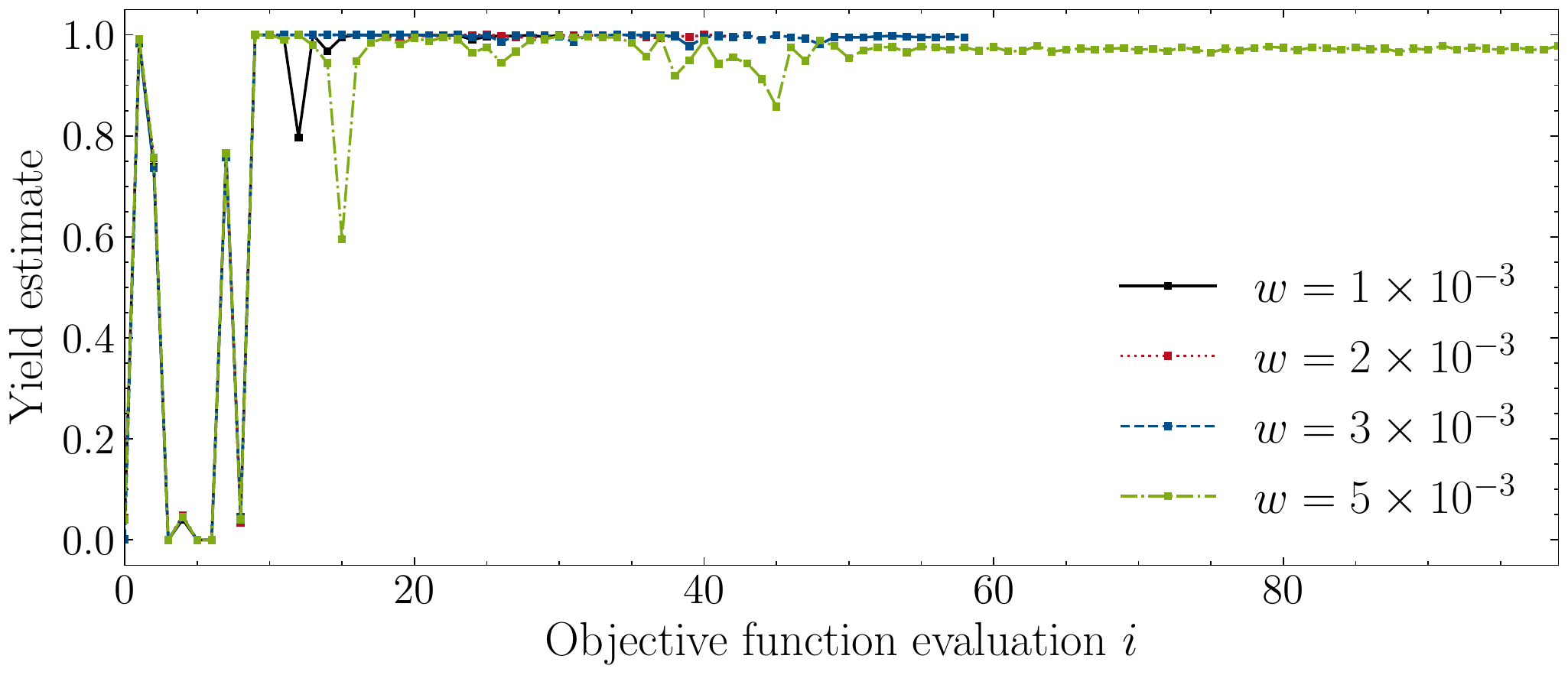}
	\caption{Propagation of the yield estimate throughout the optimization for the  weighted sum method.}
	\label{fig:ws_yield}
\end{figure}

\begin{figure}[t!]
	\centering
	\includegraphics[width=\columnwidth]{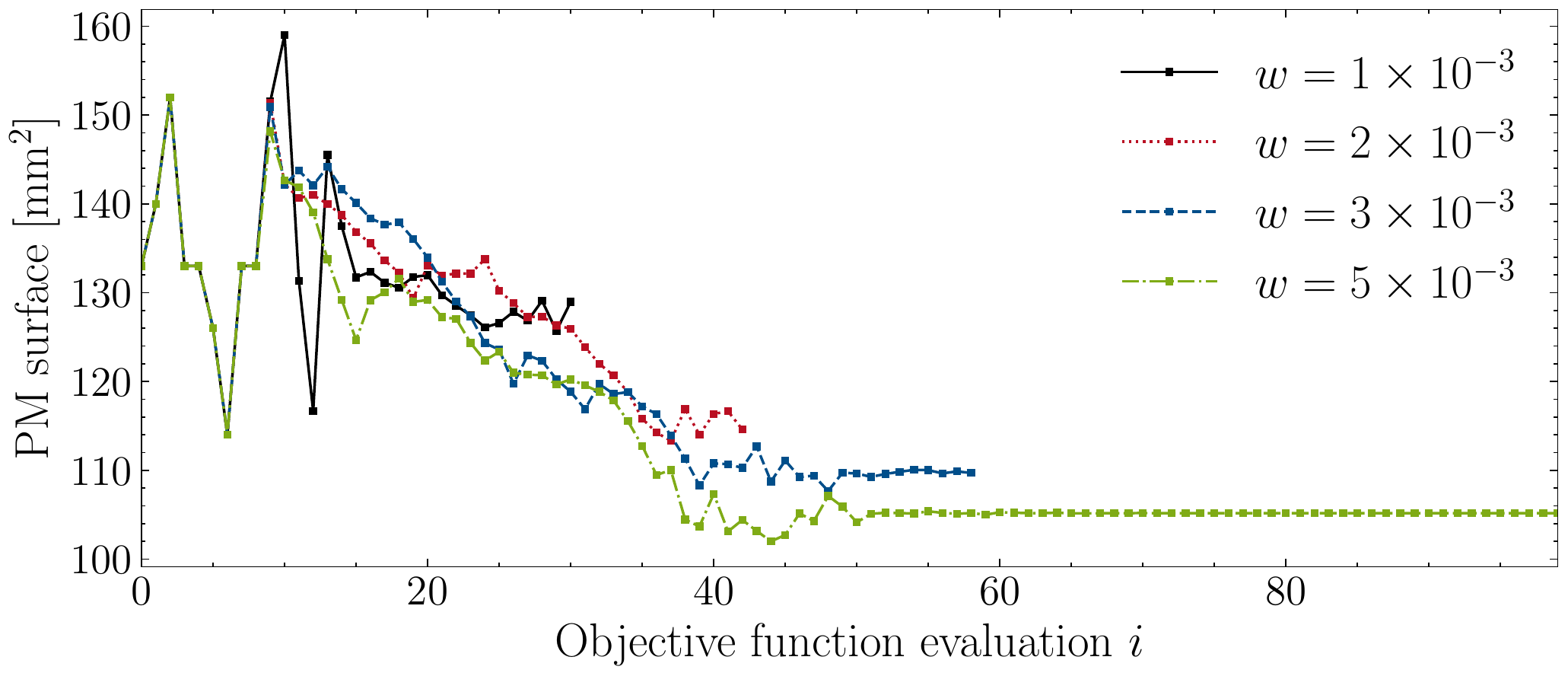}
	\caption{Propagation of the \gls{pm} surface dimensions throughout the optimization for the weighted sum  method.}
	\label{fig:ws_costs}
\end{figure}

 Furthermore, the tradeoff between the surface dimension and the yield estimate is visible in Table \ref{tab:stats_ws}, as the optimized yield estimate $ \tilde{Y}_\mathrm{MC}^{(\mathrm{opt})}$ decreases when the surface dimensions $C(\mathbf{x})$ decreases. The propagation of the yield estimate and the \gls{pm} surface dimensions is shown in Figure \ref{fig:ws_yield} and \ref{fig:ws_costs}. In comparison to the propagation of the objectives in the $\varepsilon$-constraint method, the optimization needs a larger amount of function evaluations to siginificantly decrease the surface of the \gls{pm}s, as there is no target defined for the surface dimensions as with $C_\mathrm{max}$ in the $\varepsilon$-constraint method. 

\subsubsection{Multistart procedure}
\begin{figure}[b]
	\centering
	\includegraphics[width=\columnwidth]{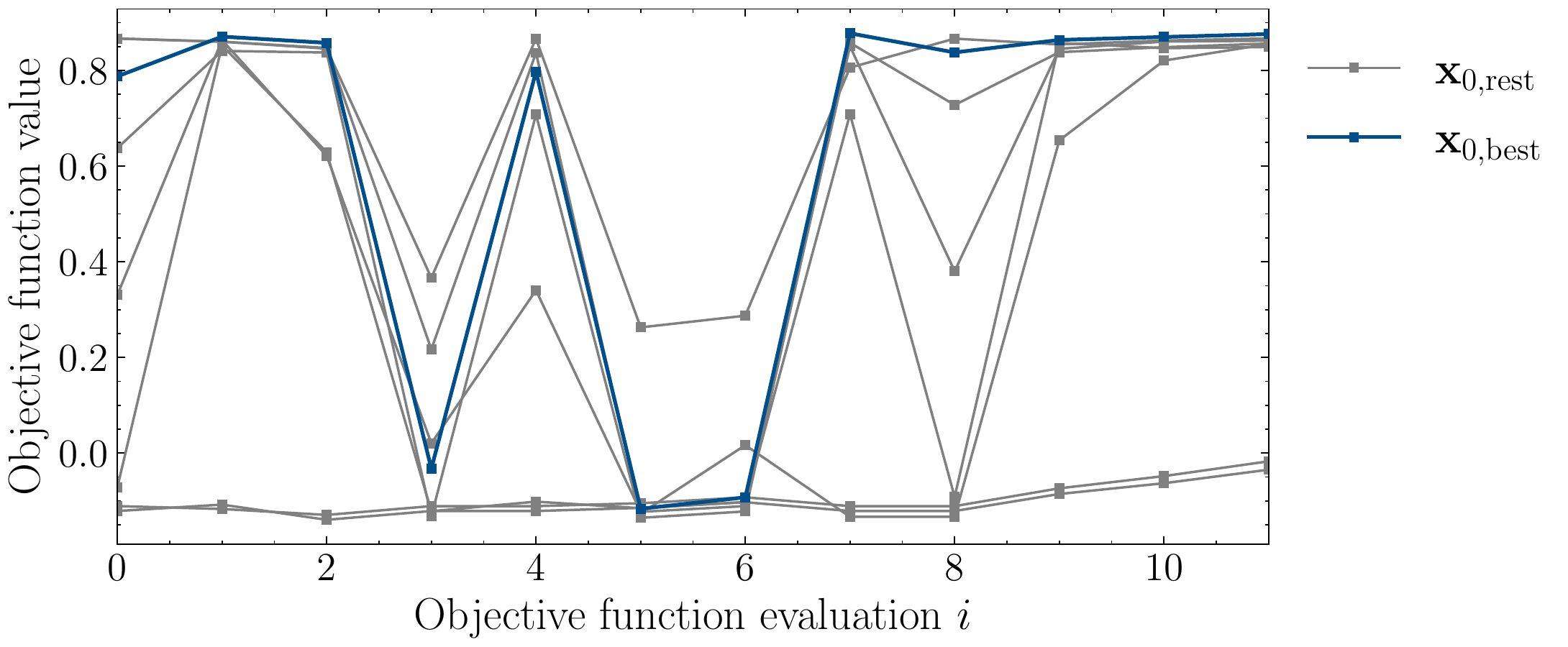}
	\caption{Demonstration of exploration phase where the best performing starting point is highlighted in blue.}
	\label{fig:multistart_procedure}
\end{figure}
\begin{figure}[b!]
	\centering
	\includegraphics[width=\columnwidth]{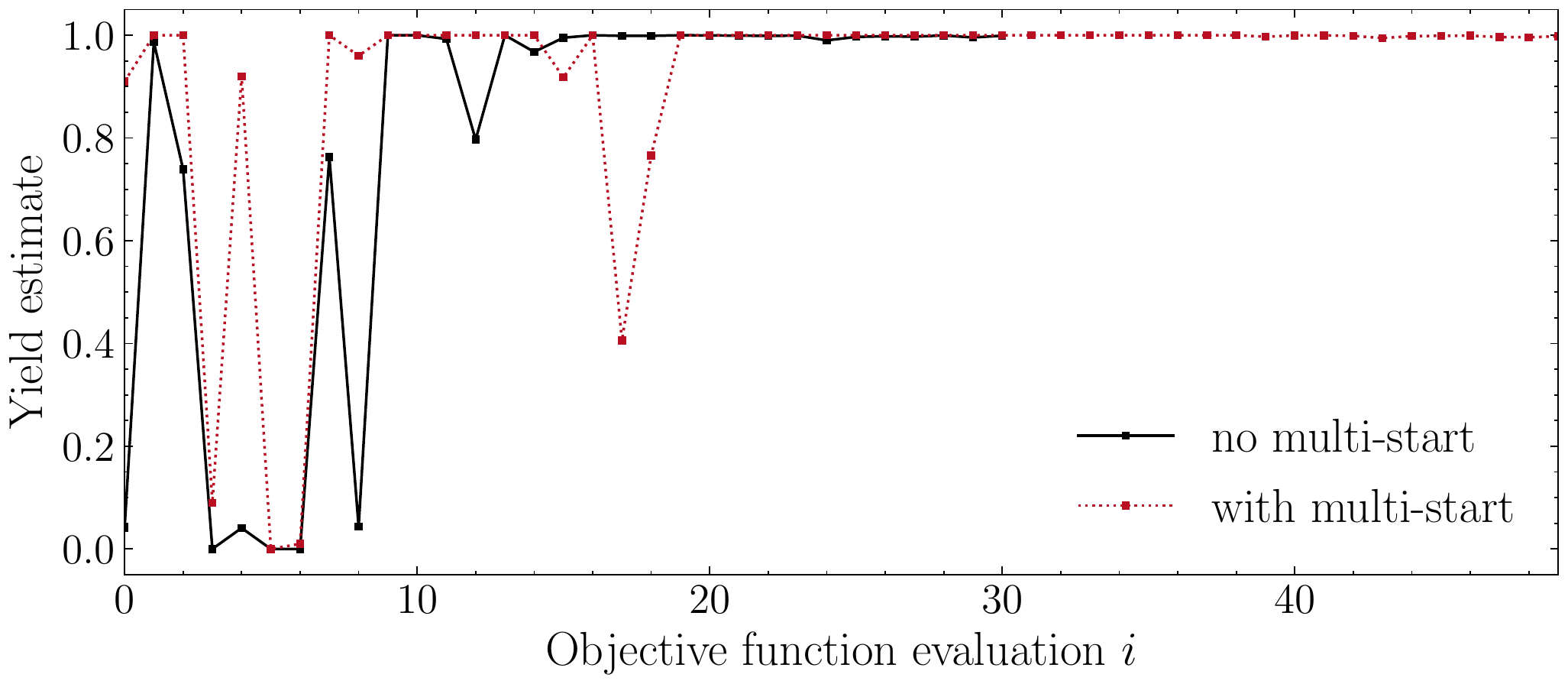}
	\caption{Comparison of the propagation of the yield estimate throughout the optimization for the  weighted sum method with and without the multi-start procedure.}
	\label{fig:multistart_procedure_compare_yield}
\end{figure}
\begin{figure}[t!]
	\centering
	\includegraphics[width=\columnwidth]{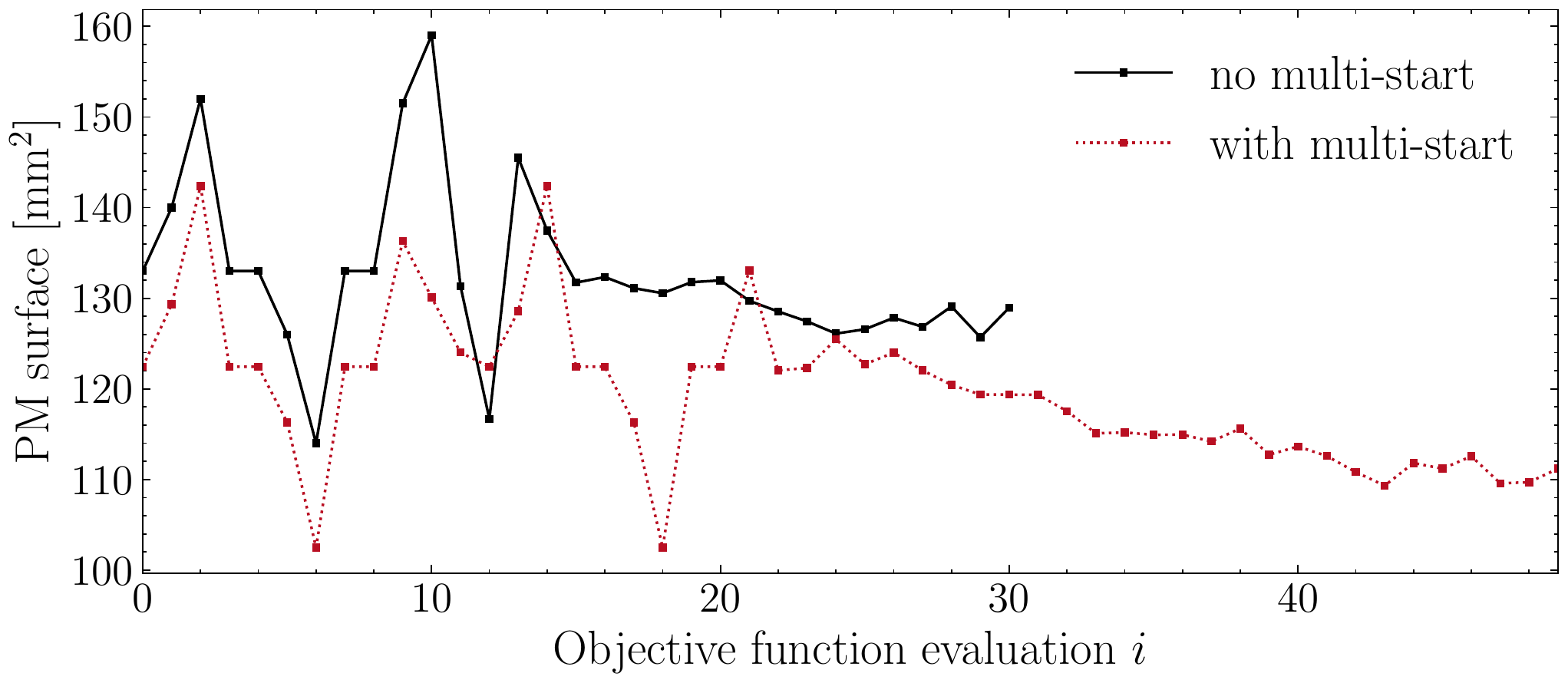}
	\caption{Comparison of the propagation of the PM surface dimensions throughout the optimization for the weighted sum method with and without the multi-start procedure.}
	\label{fig:multistart_procedure_compare_costs}
\end{figure}
\begin{figure}[t!]
	\centering
	\includegraphics[width=.5\columnwidth]{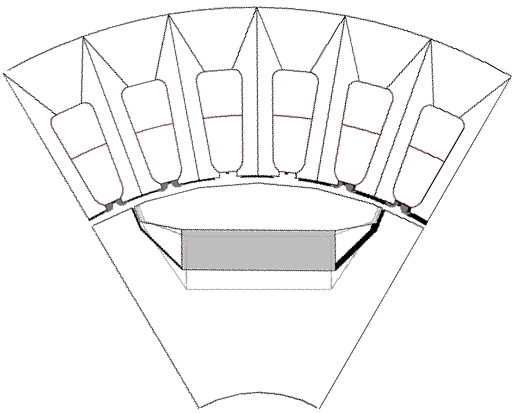}
	\caption{Optimized design (multi-start method). Here, the fille rectangle displays the optimized design and the contours the original design dimensions.}
	\label{fig:opti_design_cad}
\end{figure}

This section presents the results of the advanced method of including a multistart procedure into the weighted sum optimization to further improve the performance. In general, the two described methods of the $\varepsilon$-constraint and weighted sum method have the disadvantage that a degree of freedom needs to be set, that has a significant influence on the outcome of the optimization as shown in Table \ref{tab:stats_con} and \ref{tab:stats_ws}. Especially, the latter has difficulties with nonconvex objectives and is prone to run into suboptimal local minima. Therefore the multistart procedure is applied with the weighted sum method setting that led to higher surface dimensions of $C(\mathbf{x})=\SI{128.963}{\milli\meter\squared}$. The multistart procedure for the first twelve function evaluations with low fidelity, i.e., $N_\mathrm{MC}=100$ can be seen in Figure \ref{fig:multistart_procedure}. The starting point $\mathbf{x}_{0,\mathrm{best}}$ represents the design vector $\mathbf{x}$ that led to the best objective value after twelve function evaluations and is highlighted in blue. After the exploration phase with $N_\mathrm{MC}=100$ is completed, the exploitation phase for that starting point is continued with high-fidelity ($N_\mathrm{MC}$=2500). The outcome is shown in Table \ref{tab:stats_ws_multi}. We see that by including the multistart procedure, the \gls{pm} surface dimensions were further decreased by $\SI{14}{\percent}$. 
This shows, that the disadvantage of scalarization methods to converge into suboptimal local optima can be compensated by the multistart procedure.
Furthermore, Figures \ref{fig:multistart_procedure_compare_yield} and \ref{fig:multistart_procedure_compare_costs} show the direct comparison of the weighted sum method with and without the multistart procedure by the propagation of the yield estimate and the \gls{pm} surface dimensions. It can be seen that the setting of the weighted sum approach without the multistart procedure runs into the local optimum with surface dimensions of $\SI{128.963}{\milli\meter\squared}$, whereas the application of the multistart procedure leads to a further reduction to $C(\mathbf{d})=\SI{111.203}{\milli\meter\squared}$. An exemplary design optimization for the multistart method is shown in Figure \ref{fig:opti_design_cad}.
\begin{table}[t!]
	\begin{center}
		\caption{Computation statistics of weighted sum method with and without the multi-start procedure.}
		\label{tab:stats_ws_multi}
		\def\arraystretch{1.2}
		\begin{tabular}{l|l|l|l|l}
			\hline
			\hline
			Multi-start &$ \tilde{Y}_\mathrm{MC}^{(\mathrm{opt})}$ & $C(\mathbf{x})$ &   $n_\mathrm{fev}$ &   Total FE evaluations \\
			\hline
			no  &$0.999$ & $128.963$ & $31$ &   $20 + 259 = 279$ \\
			yes &$0.998$ &$111.203$ & $135$ &  $20 +394 = 414$ \\
			\hline  
			\hline
		\end{tabular}
	\end{center}
\end{table}

\begin{figure}[t!]
	\centering
	\includegraphics[width=\columnwidth]{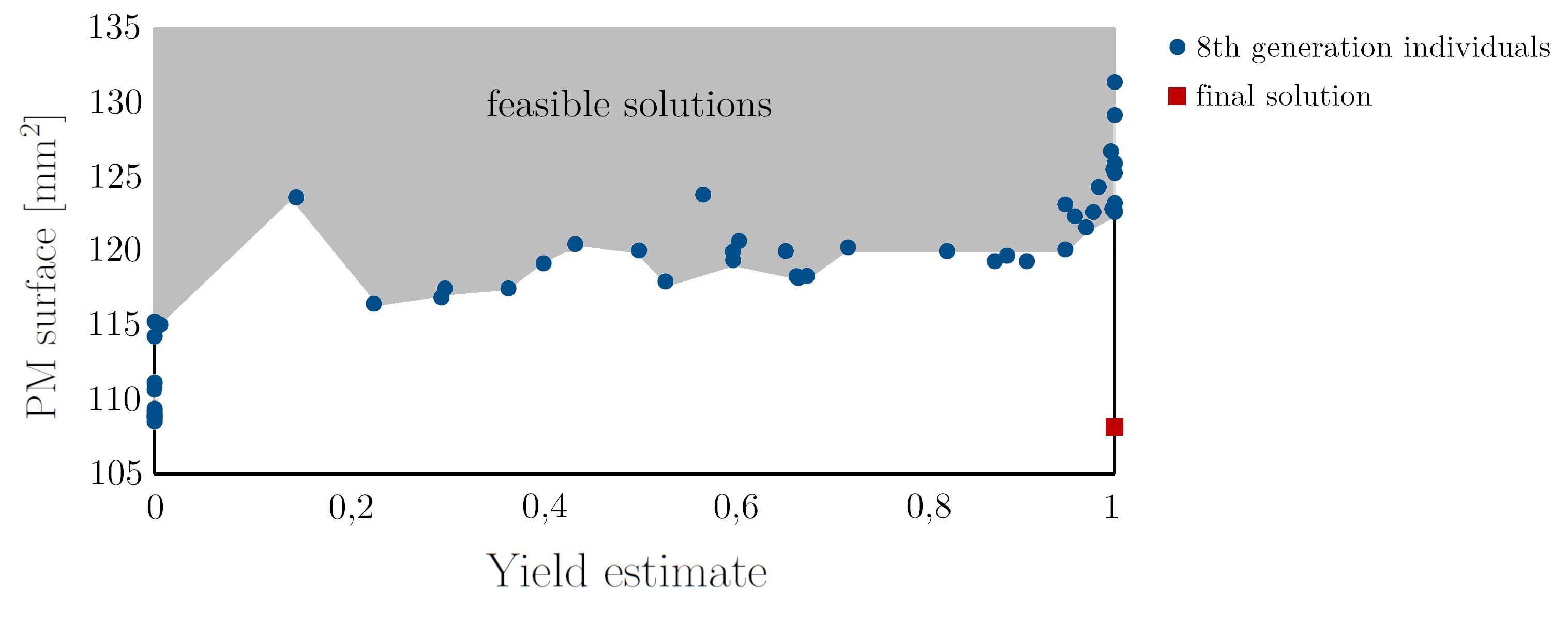}
	\caption{Visualization of the pareto front of the 8th generation of the genetic algorithm for the stated problem similar to Figure \ref{fig:ParetoFront}. NSGA-II algorithm converges to single solution marked with red square.}
	\label{fig:Results_GA_edited_final}
\end{figure} 

\subsubsection{Genetic Algorithm}
This section gives a short outlook into the application of a genetic algorithm applied to the \gls{moo} problem. Especially for this kind of computationally expensive problems a fast computation of a single function evaluation is mandatory. With the application of the \gls{gpr}-Hybrid approach it is possible to reduce the percentage of needed FE model evaluations to approx. $11000$, which corresponds to \SI{0.441}{\percent} of the total \gls{qoi} evaluations needed for the genetic algorithm simulation. For the simulation itself we set a budget of $1000$ objective function evaluations, i.e., yield estimations, a starting population size of $100$ with $50$ offsprings in future generations. A snapshot of the resulting pareto front can be seen in Figure \ref{fig:Results_GA_edited_final} that displays the individuals of the $8$th generation with the feasible solutions depicted in gray. Finally, the NSGA-II algorithm terminates the computation after $1000$ function evaluations are conducted and converges to a single point (depicted by the red square in the same figure) with a yield estimate of $1.0$ and \gls{pm} surface dimensions of \SI{108,146}{\milli \meter \squared}. The advantage of the \gls{ga} simulation is that no further parameters such as a bound $C_\mathrm{max}$ or weights $w$ need to be set accordingly to acquire good optima.

\section{Conclusion}\label{sec:Conclusion}
This paper introduces the application of the GPR-Hybrid method to electrical machines. We show that by using machine learning, the computational effort of yield estimation and yield optimization can be significantly reduced, while maintaining high accuracy.
Furthermore, for the design of the electrical machine, multi-objective optimization is applied in order to increase the reliability, i.e., the yield, and to decrease the costs, i.e., the size of the magnets.
The scalarization methods $\varepsilon$-constraint and weighted sum are compared. Both lead to overall high values for the yield estimate, while the rate of decrease of the \gls{pm} surface dimensions depends on the choice of the scalarization parameters, which is a drawback in practice.
On the other hand, the weighted sum method is extended with a multistart procedure that further improves the optimization performance of the utilized methods. Although scalarization parameters still have to be chosen, their impact on the optimization result is greatly reduced.
Additionally, a genetic algorithm (GA) is applied, which does not require such a parameter selection. Both methods, GA and weighted sum with multistart procedure, achieve to find a similar optimal solution. The multistart approach requires $4\,\%$ of the finite element (FE) model evaluations compared to \gls{ga}. 


For future research, however, multiple advanced concepts can be applied. These are, for example, the addition of multiple QoIs for the electrical machine such as the total harmonic distortion, adaptive weight selection for the weighted sum method, or adaptive numbers of the
MC sample points. 


\section*{Acknowledgments}
This work has been supported by the Graduate School CE within the Centre for Computational Engineering at Technische Universität Darmstadt, the Federal Ministry of Education and Research (BMBF) and the state of Hesse as part of the NHR Program and the SFB TRR 361 CREATOR (grant number 492661287) funded by the German Research Foundation DFG.

\vfill

\end{document}